\documentclass[a4wide,11pt]{article}

\pdfoutput=1 

\usepackage[table]{xcolor}
\usepackage{colortbl}
\RequirePackage{ifpdf} 
\usepackage{amsmath,amsfonts} 
\usepackage{mathtools}
\usepackage{cases}
\usepackage[T1]{fontenc}
\usepackage{upquote,multirow,slashed,physics}
\usepackage{authblk}
\usepackage{cite,appendix}
\usepackage{empheq}
\usepackage{lipsum}

\usepackage{jheppub}

\usepackage{pstricks}
\usepackage[final]{pdfpages} 
\usepackage{ifpdf} 
\usepackage{slashed}
\usepackage[normalem]{ulem}
\usepackage{color}
\usepackage{xcolor}
\definecolor{urlblue}{rgb}{0.2,0.4,0.7}
\definecolor{citegreen}{rgb}{0,0.6,0.2}
\definecolor{linkred}{rgb}{0.9,0.2,0.1}
\usepackage{hyperref}
\hypersetup{colorlinks=true, citecolor=citegreen, linkcolor=blue, urlcolor=blue}

\usepackage{graphics}
\usepackage[labelformat=simple]{subcaption}
\usepackage{etoolbox} 
\usepackage{fixmath}
\usepackage{psfrag}
\usepackage{autobreak}
\usepackage{marginnote}
\usepackage{scalerel}
\usepackage{enumitem}
\usepackage{appendix}
\usepackage{caption} 
\usepackage{notoccite} 

\newcommand{\NOdisplay}[1]{ }

\def\MSbar{\overline{\mathrm{MS}}}
\def\TR{{\displaystyle \mathrm{T}_{F}}}
\def\g5{\gamma_5}
\def\Zps5{Z^{ps}_5}
\def\Zpsbar{\overline{Z}^{\, ps}_5}
\def\Zpsf{Z^{ps}_f}
\def\form{{\fontfamily{qcr}\selectfont
FORM}}

\def\diagen{{\fontfamily{qcr}\selectfont
DiaGen}}
\def\idsolver{{\fontfamily{qcr}\selectfont
IdSolver}}
\def\forcer{{\fontfamily{qcr}\selectfont
Forcer}}
\def\feyngame{{\fontfamily{qcr}\selectfont
FeynGame}}

\preprint{TTK-24-42, P3H-24-076, MPP-2024-201}
\title{Renormalization of the pseudoscalar operator at four loops in QCD}

\author{Long Chen$^{a}$,}
\emailAdd{longchen@sdu.edu.cn}
\affiliation{$^{a}$School of Physics, Shandong University, Jinan, Shandong 250100, China}

\author{Micha\l{} Czakon$^{b}$,}
\emailAdd{mczakon@physik.rwth-aachen.de}
\affiliation{$^{b}$Institut f\"ur Theoretische Teilchenphysik und Kosmologie, RWTH Aachen University, Sommerfeldstra{\ss}e~16, 52056 Aachen, Germany}

\author{Marco Niggetiedt$^{c}$}
\emailAdd{marco.niggetiedt@mpp.mpg.de}
\affiliation{$^c$Max-Planck-Institut f\"ur Physik, Boltzmannstra{\ss}e~8, 85748 Garching, Germany}

\abstract{
We present the renormalization constant of the pseudoscalar operator defined with a non-anticommuting $\gamma_5$ in dimensional regularization up to four-loop order in perturbative Quantum Chromodynamics (QCD).
Furthermore, by virtue of renormalization-group invariance of the relation between the scalar and the pseudoscalar operator, we predict the $\MSbar$ factor of the renormalization constant for the latter at five-loop order in QCD.
}

\begin{document}

\allowdisplaybreaks[4]
\unitlength1cm
\maketitle
\flushbottom

\section{Introduction}
\label{sec:intro}

There are well-known technical issues related to the treatment of the chiral matrix $\g5$ in loop diagrams in Dimensional Regularization (DR)~\cite{tHooft:1972tcz,Bollini:1972ui}: 
a Dirac algebra with a fully anticommuting $\gamma_5$ in generic D$(\,\neq 4)$ dimensions prohibits a non-vanishing value of the trace of the product of one $\gamma_5$ and four $\gamma$ matrices in 4 dimensions.
In case of a non-anticommuting $\gamma_5$ scheme~\cite{tHooft:1972tcz,Akyeampong:1973xi,Breitenlohner:1977hr,Breitenlohner:1975hg,Breitenlohner:1976te}, certain symmetry properties of matrix elements or Green functions of local composite operators may not be preserved at the bare level.
As a consequence, additional anomalous terms emerge in bare
expressions of dimensionally regularized $\g5$-dependent diagrams with
ultraviolet (UV)
divergences~\cite{Bardeen:1972vi,Breitenlohner:1977hr,Breitenlohner:1975hg,Breitenlohner:1976te,Chanowitz:1979zu,Gottlieb:1979ix,Trueman:1979en,Fujii:1980yt,Bonneau:1980yb,Espriu:1982bw,Larin:1991tj,Larin:1993tq,Bos:1992nd},
necessitating their elimination order-by-order with the help of $\g5$-related symmetry-restoration counterterms.

When focusing solely on the chiral-symmetric QCD corrections to external local composite operators containing $\g5$, identifying the potential $\g5$-related symmetry-restoration renormalizations is not as intricate as it would be in a chiral gauge theory,~e.g.~the Standard Model\footnote{Maintaining the anticommutativity of $\g5$ in DR is not very straightforward either, and we refer to refs.~\cite{Bardeen:1972vi,Chanowitz:1979zu,Gottlieb:1979ix,Korner:1991sx,Kreimer:1993bh,Chen:2023lus,Chen:2024zju} for a more comprehensive discussion.}.
Non-anticommuting $\g5$ schemes are commonly utilized in practical calculations for such cases, both owing to the technical ease of implementing such a $\g5$ treatment and the availability of results for many of the necessary additional renormalization constants up to high orders in QCD~\cite{Larin:1991tj,Larin:1993tq,Larin:1997qq,Rittinger:2012the,Ahmed:2021spj,Chen:2021gxv,Chen:2022lun}. 
In particular, the complete results for the renormalizations of flavor non-singlet and singlet axial-current operators were derived in refs.~\cite{Chen:2021gxv,Chen:2022lun} up to four-loop order, and partially extended to five-loop order in QCD through an efficient use of an off-shell axial Ward-Takahashi identity~\cite{Adler:1969gk}. 
~\\

In the present study, we calculate the renormalization constant for a pseudoscalar operator regularized with a non-anticommuting $\gamma_5$ in DR~\cite{tHooft:1972tcz,Akyeampong:1973xi,Breitenlohner:1977hr,Breitenlohner:1975hg,Breitenlohner:1976te} up to four-loop order in QCD. 
We specifically employ the non-anticommuting $\gamma_5$ variant as prescribed in refs.~\cite{Larin:1991tj,Larin:1993tq}.
We further highlight that, due to the renormalization-group invariance of the appropriately defined renormalization constant for this operator (detailed in the following section~\ref{sec:tech}), its $\MSbar$ factor at five-loop order in QCD, i.e.\ the pure-pole contribution with respect to the dimensional regulator $\epsilon$ (corresponding to spacetime dimension $D = 4-2\epsilon$), can be predicted with the help of our comprehensive four-loop results. 
Indeed, by this method, we have verified that the three-loop results obtained quite some time ago in ref.~\cite{Larin:1993tq} (the version on arXiv), which we have completely reproduced independently, were adequate to determine the $\MSbar$ renormalization constant of this operator at $\mathcal{O}(\alpha_s^4)$ in QCD, in agreement with our direct computation.
This serves as a strong check of our computational framework utilized for calculating the higher-order results.

In the upcoming section~\ref{sec:tech}, we provide a summary of the preliminary knowledge on the renormalization of a pseudoscalar operator defined with a non-anticommuting $\g5$ in DR, including a discussion of a decomposition aimed at explicitly isolating the component solely due to $\g5$-related symmetry-restoration.
Our main results for the renormalization constants of interest are presented in section~\ref{sec:res}.
We conclude in section~\ref{sec:conc}.

\section{Preliminaries and Technicalities}
\label{sec:tech}

We consider the renormalization of the pseudoscalar operator $J^{ps}$, along with its scalar counterpart $J^{s}$, in QCD:
\begin{eqnarray} 
\label{eq:Jps}
\left[ J^{ps}\right]_{R} = Z^{ps} \, \bar{\psi}^{B}  \, i\,\gamma_5 \, \psi^{B}\,,
\quad\quad 
\left[ J^{s}\right]_{R} = Z^{s} \, \bar{\psi}^{B}  \, \psi^{B} \; ,
\end{eqnarray}
where $\psi^{B}$ denotes a bare quark field and the subscript $R$ at a square bracket denotes multiplicative operator renormalization.

The $\gamma_5$ matrix in the pseudoscalar operator $J^{ps}$ is treated as non-anticommuting in DR~\cite{tHooft:1972tcz,Akyeampong:1973xi,Breitenlohner:1977hr,Breitenlohner:1975hg,Breitenlohner:1976te},
according to: 
\begin{align}
\label{eq:gamma5}
\gamma_5=-\frac{i}{4!}\epsilon^{\mu\nu\rho\sigma}\gamma_{\mu}\gamma_{\nu}\gamma_{\rho}\gamma_{\sigma}\,,
\end{align}
with the Levi-Civita tensor\footnote{We use the convention $\epsilon^{0123} = -\epsilon_{0123} = +1$.} $\epsilon^{\mu\nu\rho\sigma}$ manipulated according to refs.~\cite{Larin:1991tj,Zijlstra:1992kj,Larin:1993tq}.

In case of pure QCD corrections to matrix elements of the pseudoscalar operator with massless quarks, the so-called singlet Feynman diagrams vanish. 
These are those diagrams where the pseudoscalar vertex attaches to a closed fermion loop. 
The vanishing is due to the presence of an odd number of $\gamma$-matrices in Dirac traces. 
In consequence, non-vanishing contributions require the pseudoscalar vertex to attach to an open fermion line. 
Hence, bare two-point Green's functions of $J^{ps}$ are related to those of $J^{s}$ if $\g5$ is anticommuting. 
In our case of non-anticommuting $\g5$, we require the same relation to hold as part of the definition of the renormalized pseudoscalar operator:
\begin{equation} \label{eq:rencon}
  \ev{\hat{\mathrm{T}} \big[ \psi \left[ J^{ps}\right]_{R} \bar{\psi} \big]}{\Omega} = - \ev{\hat{\mathrm{T}} \big[ \psi \left[ J^{s}\right]_{R} \bar{\psi} \big]}{\Omega} \gamma_5 \;, 
\end{equation}
where $\hat{\mathrm{T}}\big[\;\big]$ denotes the time-ordering operation and $|\Omega\rangle$ is the vacuum state.

If not for the mentioned $\g5$ issue in DR, the renormalization constant $Z^{ps}$ would be identical to $Z^{s}$, the latter equal to the mass renormalization constant in e.g.~the $\MSbar$ scheme. 
Since the anticommutativity of $\g5$ is spoiled at the bare level in the  regularization scheme adopted in this work, these two quantities are no longer the same in general. 
To have this point manifested more clearly, we find it useful to introduce the following factor: 
\begin{equation} \label{eq:zps5def}
\Zps5 \equiv \frac{Z^{ps}}{Z^s} =  \Zpsf\, \Zpsbar
\end{equation}
Although this is not strictly necessary, we assume $Z^s$ to be determined in the $\MSbar$ scheme.
In the second equality in \eqref{eq:zps5def}, this factor $\Zps5$ is subsequently parameterized as a product $\Zpsbar \, \Zpsf$ where $\Zpsbar$ contains only poles in the dimensional regulator $\epsilon$, whereas $\Zpsf$ has no poles in $\epsilon$ and is truncated to $\mathcal{O}(\epsilon^0)$.
The finite $\Zpsf$ is introduced on top of $\Zpsbar$ to ensure that eq.~\eqref{eq:rencon} is fulfilled\footnote{This factor should not be confused with a factor that could be introduced to transform from the $\MSbar$ renormalization scheme for the (pseudo)scalar operator or Yukawa coupling, to another renormalization scheme, such as the on-shell scheme where the renormalization is associated with the on-shell mass renormalization.}.

It follows from the renormalisation condition \eqref{eq:rencon} that $\Zps5$ defined in \eqref{eq:zps5def} is renormalization-group invariant, meaning that its anomalous dimension vanishes in the 4-dimensional limit:
\begin{equation} \label{eq:AMDofZps5}
\mu^2\frac{\mathrm{d} \ln \Zps5}{\mathrm{d} \mu^2}\Big|_{\epsilon = 0}  = 0 \,,
\end{equation}
with $\mu$ the auxiliary scale introduced in DR. 
As a consequence of the renormalization-group invariance \eqref{eq:AMDofZps5} we then have 
\begin{eqnarray} \label{eq:AMDofZms5} 
\mu^2\frac{\mathrm{d} \ln \Zpsbar}{\mathrm{d} \mu^2} 
&=& - \mu^2\frac{\mathrm{d} \ln \Zpsf}{\mathrm{d} \mu^2}\Big|_{\epsilon = 0}  \,
= - \frac{\mathrm{d} \ln \Zpsf}{\mathrm{d} \ln \alpha_s} \big( - \epsilon + \beta \big)|_{\epsilon = 0} \nonumber\\
&=& -\beta\, \frac{\mathrm{d} \ln \Zpsf}{\mathrm{d} \ln \alpha_s} 
\,,
\end{eqnarray}
where the QCD $\beta$ function is defined as the anomalous dimension of the $\MSbar$-renormalized $\alpha_s$ via $\mathrm{d} \ln \alpha_s / \mathrm{d} \ln \mu^2 = -\epsilon + \beta \,$.
In the last equality in \eqref{eq:AMDofZms5}, we have also used the fact that the finite $\Zpsf$ has, by definition, no explicit dependence on $\epsilon$.
Since the perturbative expansion of so-defined $\beta$ starts from $\mathcal{O}(\alpha_s)$, \eqref{eq:AMDofZms5} implies that the result for the anomalous dimension of the $\MSbar$ renormalization constant $\Zpsbar$ at $\mathcal{O}(\alpha_s^{N+1})$ requires merely the knowledge of $\Zpsf$ at $\mathcal{O}(\alpha_s^{N})$.
The pure $\MSbar$-renormalization constant $\Zpsbar$ can then be uniquely reconstructed from its anomalous dimension given by the r.h.s.~of \eqref{eq:AMDofZms5}.
This equation \eqref{eq:AMDofZms5} also helps to appreciate
technically why the perturbative correction to $\Zpsbar$ starts from
$\mathcal{O}(\alpha_s^2)$ with a simple pole in $\epsilon$, resembling
(though not identical to) that of the non-anomalous axial-vector
current. 
Notice that this structure is not observed in the complete renormalisation constant $Z^{ps}$ whose concrete expression depends on the renormalization condition in use.
~\\

\begin{figure}[htbp]
  \centering
  \begin{subfigure}[b]{0.25\textwidth}
    \centering
    \includegraphics[scale=0.18]{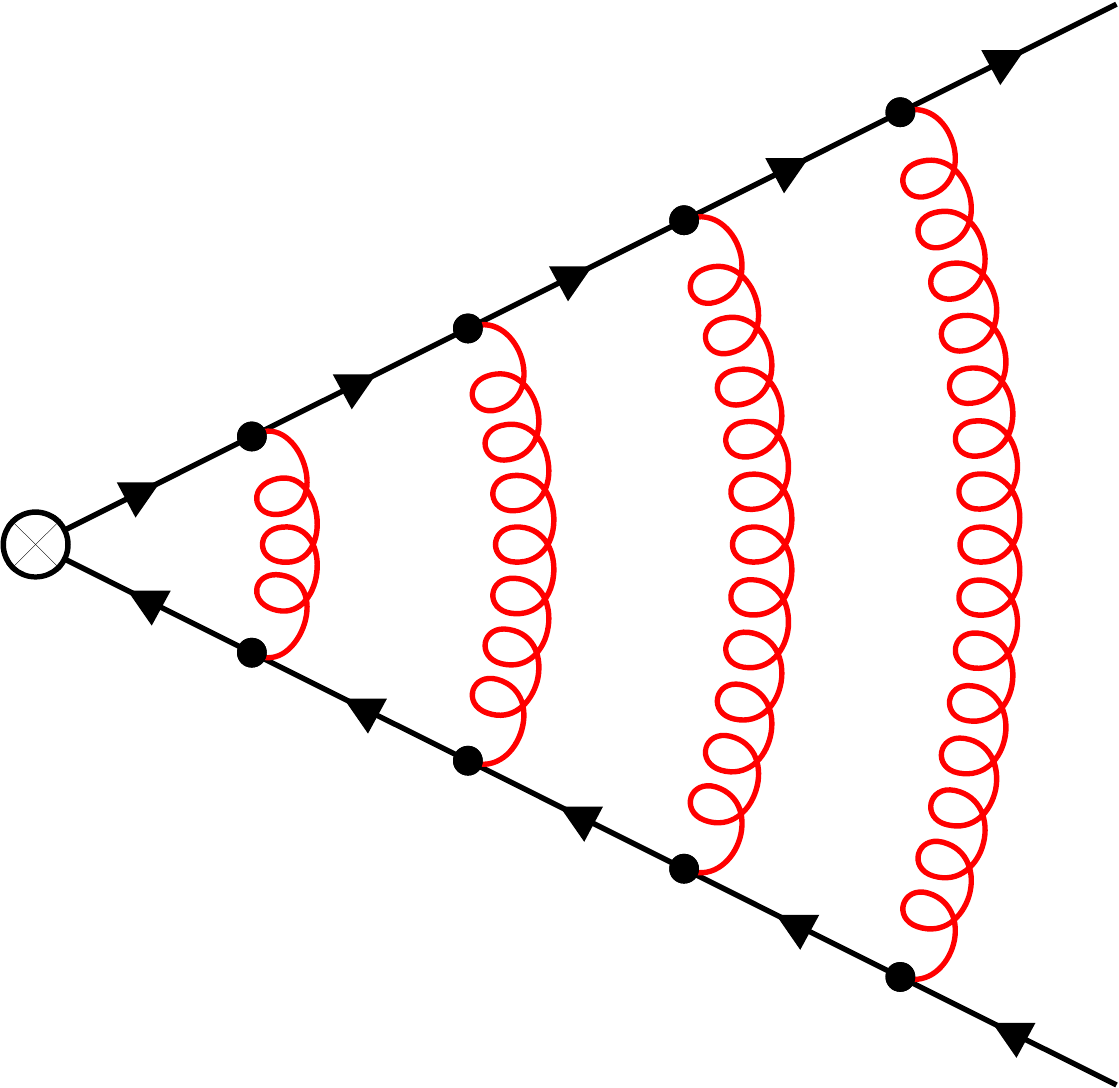}
  \end{subfigure}%
  \begin{subfigure}[b]{0.25\textwidth}
    \centering
    \includegraphics[scale=0.18]{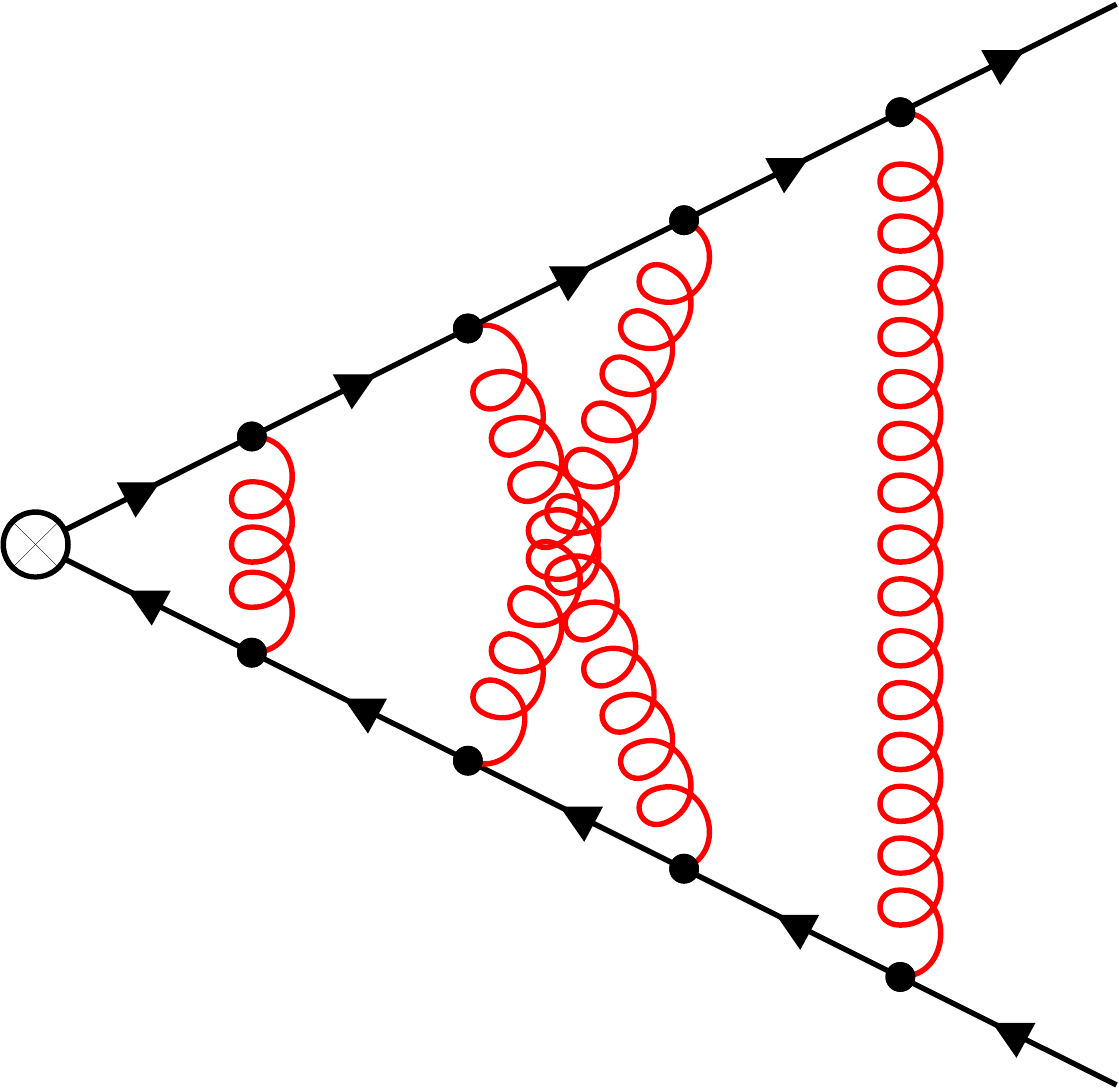}
  \end{subfigure}%
  \begin{subfigure}[b]{0.25\textwidth}
    \centering
    \includegraphics[scale=0.18]{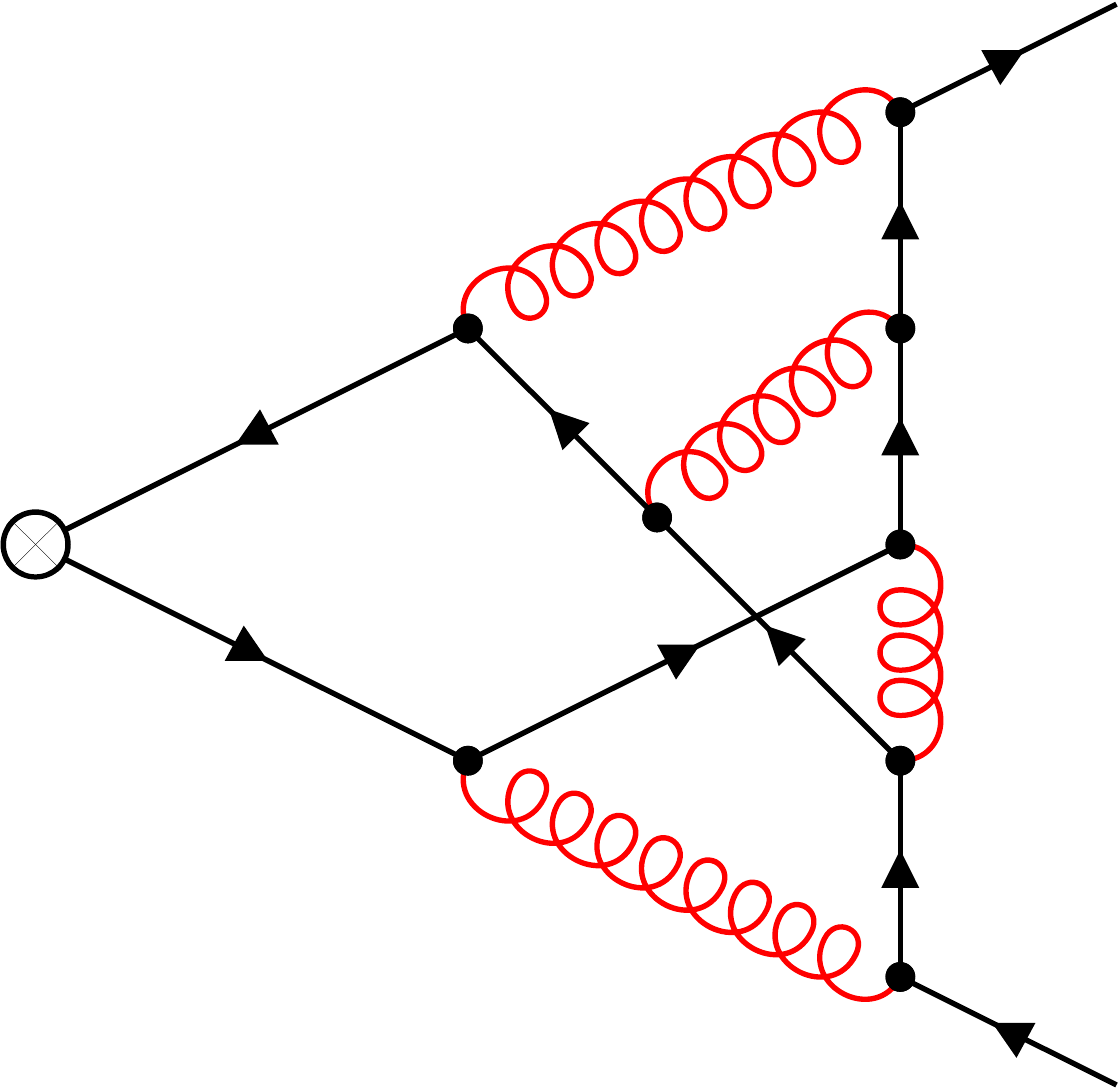}
  \end{subfigure}%
  \begin{subfigure}[b]{0.25\textwidth}
    \centering
    \includegraphics[scale=0.18]{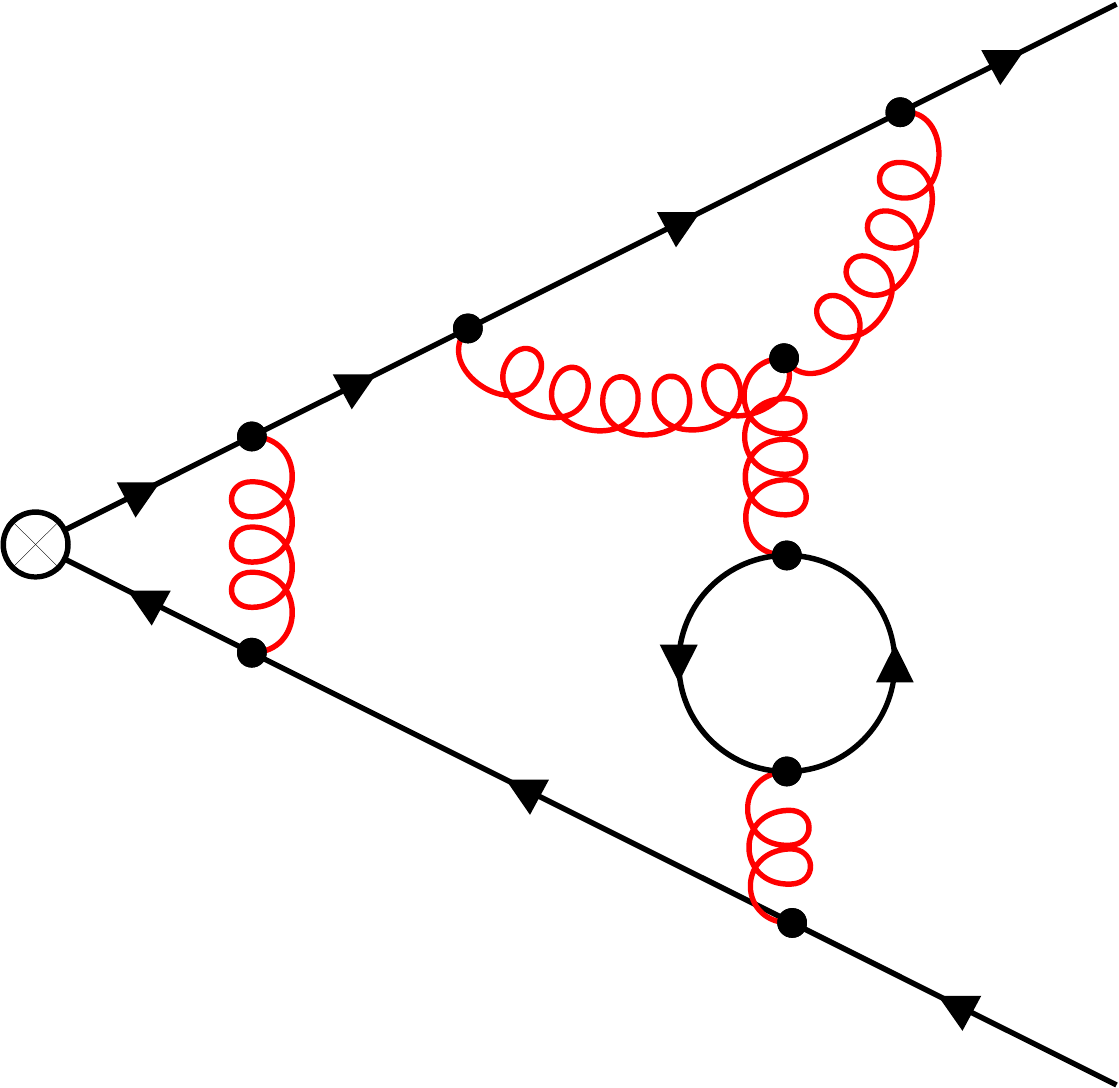}
  \end{subfigure}%
  \caption{Representative Feynman diagrams at four-loop order. The encircled cross indicates  the insertion of the external pseudo-scalar operator. The solid lines with arrows represent quark propagators and the circular lines in red color denote gluon propagators.}
    \label{fig:diags}
\end{figure}

The explicit results for $\Zpsbar$ and $\Zpsf$, as well as $Z^{s}$, are extracted from the perturbative expressions for the matrix elements of the pseudoscalar and scalar operator defined in \eqref{eq:Jps} between the vacuum and a pair of external quarks, up to four-loop order. 
To simplify the calculation and also get rid of divergences not of UV origin, we take a kinematic setup where the momentum flow through the pseudoscalar operator is zero and both external quarks are set off-shell.
The pertubative QCD corrections to these matrix elements are computed in terms of Feynman diagrams, which are subsequently manipulated in a similar fashion to our previous calculations~\cite{Ahmed:2021spj,Chen:2021gxv,Chen:2022lun}.
More specifically, symbolic expressions of the contributing Feynman diagrams to four-loop order are generated by the diagram generator~\diagen~\cite{diagen}.\footnote{The C++ library \diagen~provides, besides diagram generation for arbitrary Feynman rules, topological analysis tools and an interface to the C++ library \idsolver~that allows to directly apply integration-by-parts identities to the integrals occurring in the generated diagrams. \idsolver~has been originally written for the calculation of ref.~\cite{Czakon:2004bu}, while \diagen~predates this software.
}
There are in total around six thousand diagrams generated for the
needed matrix elements.
A few representative Feynman diagrams at four-loop order are shown in figure~\ref{fig:diags}, drawn using \feyngame~\cite{Harlander:2020cyh,Harlander:2024qbn}.
Feynman-rules substitution, the color algebra with a generic SU($N_c$) group and $D$-dimensional Lorentz as well as the Dirac algebra are performed using \form~\cite{Vermaseren:2000nd}. 
The computations are done in a general Lorentz-covariant gauge for the gluon field, with the general-covariant-gauge fixing parameter $\xi$ defined through the gluon propagator $\frac{i}{k^2} \left(-g^{\mu\nu} + \xi \,\frac{k^{\mu} k^{\mu}}{k^2} \right)$, $k$ being the momentum of the gluon.
Retaining the $\xi$-dependence in the off-shell matrix element expressions up to at least three-loop order not only validates the $\xi$-independence of the obtained renormalization constants but is also essential for completing the UV renormalization at the four-loop order~\cite{Chen:2021gxv}. 
In comparison to our previous treatment of the axial-current cases~\cite{Chen:2021gxv,Chen:2022lun}, the increase in the number of $\gamma$-matrices in Dirac traces leads to a considerable rise in computational complexity: numerous four-loop planar and non-planar diagrams feature  24 $\gamma$-matrices within a single Dirac trace.

The reduction of all propagator-type loop integrals in the matrix elements, totaling around $3 \times 10^4$, to master integrals is done by means of integration-by-parts identities~\cite{Tkachov:1981wb,Chetyrkin:1981qh}, carried out efficiently using the program \forcer~\cite{Ruijl:2017cxj}. 
The analytical expressions for massless propagator-type master integrals computed in refs.~\cite{Baikov:2010hf,Lee:2011jt} are then employed to derive the explicit results for the aforementioned matrix elements.
With the help of the known four-loop renormalization of the QCD Lagrangian~\cite{vanRitbergen:1997va,Chetyrkin:2004mf,Czakon:2004bu}, 
the results for the relevant renormalization constants can be extracted.

\section{Results}
\label{sec:res}

Below we present our final perturbative results for the renormalization constants  defined in \eqref{eq:zps5def} for the pseudoscalar operator at the four-loop order in QCD with $n_f$ massless quarks. 
The perturbative QCD coupling $a_s \equiv \frac{\alpha_s}{4 \pi}$ is renormalized in the $\MSbar$ scheme.

The result for the $\MSbar$ part reads: %
\begin{eqnarray}
\label{eq:zps5_ms}
\Zpsbar &=& 1 +  a_s^2\, \Big\{  
C_A C_F \, \left(\frac{44}{3 \epsilon }\right) \,+\, C_F n_f \, \left(-\frac{8}{3 \epsilon }\right)
\Big \} \nonumber\\
&+& a_s^3\, \Big\{
C_A^2 C_F \, \left(\frac{2404}{81 \epsilon }-\frac{968}{27 \epsilon ^2}\right)\,+\, C_A C_F^2 \, \left(\frac{704}{9 \epsilon }\right) \,+\, C_A C_F n_f \, \left(\frac{352}{27 \epsilon ^2}-\frac{800}{81 \epsilon }\right) \nonumber\\
&+& C_F^2 n_f \, \left(-\frac{176}{9 \epsilon }\right)\,+\,C_F n_f^2 \, \left(\frac{16}{81 \epsilon }-\frac{32}{27 \epsilon ^2}\right)
\Big \} \nonumber\\
&+& a_s^4\, \Big\{
C_A^3 C_F \, \left(-\frac{13343}{81 \epsilon ^2}+\frac{2662}{27 \epsilon ^3}+\frac{572 \zeta _3+\frac{10915}{54}}{\epsilon }\right)
\,+\, 
C_A^2 C_F^2 \, \left(\frac{\frac{6964}{27}-1672 \zeta _3}{\epsilon }-\frac{968}{9 \epsilon ^2}\right)
\nonumber\\
&+&
C_A^2 C_F n_f \, \left(\frac{734}{9 \epsilon ^2}-\frac{484}{9 \epsilon ^3}+\frac{-\frac{488 \zeta _3}{3}-\frac{8210}{81}}{\epsilon }\right)
\,+\,
C_A C_F^3 \, \left(\frac{1056 \zeta _3+\frac{572}{3}}{\epsilon }\right)
\nonumber\\
&+&
C_A C_F^2 n_f \, \left(\frac{616}{9 \epsilon ^2}+\frac{\frac{1088 \zeta _3}{3}-\frac{1094}{27}}{\epsilon }\right)
\,+\,
C_A C_F n_f^2 \, \left(-\frac{268}{27 \epsilon ^2}+\frac{88}{9 \epsilon ^3}+\frac{\frac{32 \zeta _3}{3}+\frac{439}{81}}{\epsilon }\right)
\nonumber\\
&+&
C_F^3 n_f \, \left(\frac{-192 \zeta _3-\frac{194}{3}}{\epsilon }\right)
\,+\,
C_F^2 n_f^2 \, \left(\frac{-\frac{32 \zeta _3}{3}-\frac{164}{27}}{\epsilon }-\frac{80}{9 \epsilon ^2}\right)
\nonumber\\
&+&
C_F n_f^3 \, \left(\frac{8}{81 \epsilon ^2}-\frac{16}{27 \epsilon ^3}+\frac{52}{81 \epsilon }\right)
\Big \} \,+\, 
\mathcal{O}(a_s^5)\,.
\end{eqnarray}
The definition of the quadratic Casimir color constants is as usual: $C_A = N_c \,, \, C_F = (N_c^2 - 1)/(2 N_c) \,$ with $N_c =3$ in QCD and the color-trace normalization factor $\TR = {1}/{2}$ inserted.
The result for the finite part reads:
\begin{eqnarray}
\label{eq:zps5_f}
\Zpsf &=& 1 \,+\, a_s\, C_F \, \left(-8\right) 
\,+\, a_s^2 \, \Big \{ C_A C_F \, \left(\frac{2}{9}\right)
\,+\,
C_F n_f \, \left(\frac{4}{9}\right) \Big \}
\nonumber\\
&+& 
a_s^3 \,
\Big \{ C_A^2 C_F \, \left(-208 \zeta _3-\frac{958}{27}\right)
\,+\,
C_A C_F^2 \, \left(608 \zeta _3-\frac{800}{27}\right)
\,+\,
C_A C_F n_f \, \left(\frac{64 \zeta _3}{3}+\frac{856}{81}\right)
\nonumber\\
&+&
C_F^3 \, \left(\frac{304}{3}-384 \zeta _3\right)
\,+\,
C_F^2 n_f \, \left(-\frac{64 \zeta _3}{3}-\frac{580}{27}\right)
\,+\,
C_F n_f^2 \, \left(\frac{104}{81}\right) 
\Big \}
\nonumber\\
&+& 
a_s^4 \,
\Big \{
C_A^3 C_F \, \left(-\frac{45136 \zeta _3}{27}+\frac{49060 \zeta _5}{3}+\frac{143 \pi ^4}{15}-\frac{291659}{324}\right)
\nonumber\\
&+&
C_A^2 C_F^2 \, \left(\frac{31376 \zeta _3}{3}-\frac{195100 \zeta _5}{3}-\frac{418 \pi ^4}{15}+\frac{125920}{81}\right)
\nonumber\\
&+&
C_A^2 C_F n_f \, \left(-\frac{380 \zeta _3}{9}-\frac{400 \zeta _5}{3}-\frac{122 \pi ^4}{45}+\frac{35641}{162}\right)
\nonumber\\
&+&
C_A C_F^3 \, \left(-\frac{44648 \zeta _3}{3}+81560 \zeta _5+\frac{88 \pi ^4}{5}+\frac{17524}{27}\right)
\nonumber\\
&+&
C_A C_F^2 n_f \, \left(\frac{6952 \zeta _3}{9}-\frac{80 \zeta _5}{3}+\frac{272 \pi ^4}{45}-\frac{27362}{81}\right)
\,+\,
C_A C_F n_f^2 \, \left(-\frac{176 \zeta _3}{9}+\frac{8 \pi ^4}{45}-\frac{187}{27}\right)
\nonumber\\
&+&
C_F^4 \, \left(1312 \zeta _3-27200 \zeta _5-\frac{2068}{3}\right)
\,+\,
C_F^3 n_f \, \left(-\frac{2248 \zeta _3}{3}+160 \zeta _5-\frac{16 \pi ^4}{5}+\frac{407}{27}\right)
\nonumber\\
&+&
C_F^2 n_f^2 \, \left(\frac{176 \zeta _3}{9}-\frac{8 \pi ^4}{45}-\frac{134}{81}\right)
\,+\,
C_F n_f^3 \, \left(\frac{10}{9}-\frac{32 \zeta _3}{27}\right)
\nonumber\\
&+&
C_1 C_F n_f \, \left(1024 \zeta _3+\frac{704}{3}\right)
\,+\,
C_2 C_F \, \left(-10592 \zeta _3-21120 \zeta _5+\frac{2048}{3}\right)
\Big \}
\,+\, \mathcal{O}(a_s^5)\,,
\end{eqnarray}
where the additional color constants are defined in terms of symmetric color tensors\footnote{The symmetric tensor $d_F^{abcd}$ is defined by the color trace $\frac{1}{6} \mbox{Tr}$~$\big( T^a T^b T^c T^d + T^a T^b T^d T^c + T^a T^c T^b T^d + T^a T^c T^d T^b + T^a T^d T^b T^c + T^a T^d T^c T^b \big)$ with $T^a$ the generators of the fundamental representation of the SU($N_c$) group, and similarly $d_A^{abcd}$ for the adjoint representation.} as %
\begin{eqnarray}\label{eq:C1C2}
C_1 &\equiv& \frac{d_F^{abcd} d_F^{abcd}}{N_c^2-1} = \frac{N_c^4-6 N_c^2+18}{96 N_c^2} 
\,,\,\nonumber\\
C_2 &\equiv& \frac{d_F^{abcd} d_A^{abcd}}{N_c^2-1} = \frac{N_c \left(N_c^2+6\right)}{48}\,.
\end{eqnarray}
Up to three-loop order, our results \eqref{eq:zps5_ms} (multiplied by the renormalization constant $Z^s$ for the scalar operator) and \eqref{eq:zps5_f} agree with those given in ref.~\cite{Larin:1993tq}.
The four-loop order components of \eqref{eq:zps5_ms} and \eqref{eq:zps5_f} are novel contributions of this work.
As a by-product of our calculations, the agreement between the four-loop QCD result for $Z^s$ and the $\MSbar$-mass renormalization constant in refs.~\cite{Chetyrkin:1997dh,Vermaseren:1997fq} serves as an independent check of the latter.

The anomalous dimension of $\Zpsbar$ at $\mathcal{O}(\alpha_s^5)$ in the 4-dimensional limit does not receive contributions from the (unknown) $\mathcal{O}(\alpha_s^5)$ result for $\Zpsf$.
According to the r.h.s.~of \eqref{eq:AMDofZms5} using our full four-loop results, we obtain 
\begin{eqnarray}
\label{eq:zps5_ms_AMD}
\mu^2\frac{\mathrm{d} \ln \Zpsbar}{\mathrm{d} \mu^2}  &=&
a_s^2\, 
\Big\{
C_A C_F \, \left(-\frac{88}{3}\right)
\,+\,
C_F n_f \, \left(\frac{16}{3}\right)
\Big\}
\nonumber\\
&+& 
a_s^3\, 
\Big\{
C_A^2 C_F \, \left(-\frac{2404}{27}\right)
\,+\,
C_A C_F^2 \, \left(-\frac{704}{3}\right)
\nonumber\\
&+&
C_A C_F n_f \, \left(\frac{800}{27}\right)
\,+\,
C_F^2 n_f \, \left(\frac{176}{3}\right)
\,+\,
C_F n_f^2 \, \left(-\frac{16}{27}\right)
\Big\}
\nonumber\\
&+&
a_s^4\, 
\Big\{
C_A^3 C_F \, \left(-2288 \zeta _3-\frac{21830}{27}\right)
\,+\,
C_A^2 C_F^2 \, \left(6688 \zeta _3-\frac{27856}{27}\right)
\nonumber\\
&+&
C_A^2 C_F n_f \, \left(\frac{1952 \zeta _3}{3}+\frac{32840}{81}\right)
\,+\,
C_A C_F^3 \, \left(-4224 \zeta _3-\frac{2288}{3}\right)
\nonumber\\
&+&
C_A C_F^2 n_f \, \left(\frac{4376}{27}-\frac{4352 \zeta _3}{3}\right)
\,+\,
C_A C_F n_f^2 \, \left(-\frac{128 \zeta _3}{3}-\frac{1756}{81}\right)
\nonumber\\
&+&
C_F^3 n_f \, \left(768 \zeta _3+\frac{776}{3}\right)
\,+\,
C_F^2 n_f^2 \, \left(\frac{128 \zeta _3}{3}+\frac{656}{27}\right)
\,+\,
C_F n_f^3 \, \left(-\frac{208}{81}\right)
\Big\}
\nonumber\\
&+&
a_s^5\, 
\Big\{
C_A^4 C_F \, \left(-\frac{2555648 \zeta _3}{81}+\frac{2158640 \zeta _5}{9}+\frac{6292 \pi ^4}{45}-\frac{4098295}{243}\right)
\nonumber\\
&+&
C_A^3 C_F^2 \, \left(\frac{1346944 \zeta _3}{9}-\frac{8584400 \zeta _5}{9}-\frac{18392 \pi ^4}{45}+\frac{3475816}{243}\right)
\nonumber\\
&+&
C_A^3 C_F n_f \, \left(\frac{523472 \zeta _3}{81}-\frac{410080 \zeta _5}{9}-\frac{1760 \pi ^4}{27}+\frac{2019266}{243}\right)
\nonumber\\
&+&
C_A^2 C_F^3 \, \left(-\frac{1439968 \zeta _3}{9}+\frac{3588640 \zeta _5}{3}+\frac{3872 \pi ^4}{15}+\frac{105136}{27}\right)
\nonumber\\
&+&
C_A^2 C_F^2 n_f \, \left(-\frac{401920 \zeta _3}{27}+\frac{1557280 \zeta _5}{9}+\frac{4400 \pi ^4}{27}-\frac{1377070}{243}\right)
\nonumber\\
&+&
C_A^2 C_F n_f^2 \, \left(-\frac{1312 \zeta _3}{3}+\frac{3200 \zeta _5}{9}+\frac{1328 \pi ^4}{135}-\frac{235402}{243}\right)
\nonumber\\
&+&
C_A C_F^4 \, \left(-\frac{77440 \zeta _3}{3}-\frac{1196800 \zeta _5}{3}-\frac{119152}{9}\right)
\nonumber\\
&+&
C_A C_F^3 n_f \, \left(\frac{39776 \zeta _3}{3}-\frac{645440 \zeta _5}{3}-\frac{1408 \pi ^4}{15}-\frac{7256}{9}\right)
\nonumber\\
&+&
C_A C_F^2 n_f^2 \, \left(-\frac{60544 \zeta _3}{27}+\frac{640 \zeta _5}{9}-\frac{2528 \pi ^4}{135}+\frac{163304}{243}\right)
\nonumber\\
&+&
C_A C_F n_f^3 \, \left(\frac{2816 \zeta _3}{81}-\frac{64 \pi ^4}{135}+\frac{580}{27}\right)
\,+\,
C_1 C_A C_F n_f \, \left(\frac{45056 \zeta _3}{3}+\frac{30976}{9}\right)
\nonumber\\
&+&
C_2 C_A C_F \, \left(-\frac{466048 \zeta _3}{3}-309760 \zeta _5+\frac{90112}{9}\right)
\,+\,
C_F^4 n_f \, \left(\frac{20992 \zeta _3}{3}+\frac{217600 \zeta _5}{3}+\frac{23176}{9}\right)
\nonumber\\
&+&
C_F^3 n_f^2 \, \left(\frac{24640 \zeta _3}{9}-\frac{1280 \zeta _5}{3}+\frac{128 \pi ^4}{15}+\frac{7352}{27}\right)
\,+\,
C_F^2 n_f^3 \, \left(-\frac{1408 \zeta _3}{27}+\frac{64 \pi ^4}{135}-\frac{8360}{243}\right)
\nonumber\\
&+&
C_F n_f^4 \, \left(\frac{256 \zeta _3}{81}-\frac{80}{27}\right)
\,+\,
C_1 C_F n_f^2 \, \left(-4096 \zeta _3\right)
\,+\,
C_2 C_F n_f \, \left(\frac{98048 \zeta _3}{3}+56320 \zeta _5-\frac{20480}{9}\right)
\nonumber\\
&+&
C_3 C_F \, \left(\frac{640}{9}-\frac{5632 \zeta _3}{3}\right)
\Big\}
\,+\,
\mathcal{O}(a_s^6)\,,
\end{eqnarray} 
where there appear color constants $C_1$ and $C_2$ defined in \eqref{eq:C1C2} as well as a new one 
$$C_3 \equiv \frac{d_A^{abcd} d_A^{abcd}}{N_c^2-1} = \frac{N_c^2 \left(N_c^2+36\right)}{24}\,.$$ 
The pure $\MSbar$-renormalization constant $\Zpsbar$ at $\mathcal{O}(\alpha_s^5)$ can then be uniquely reconstructed from \eqref{eq:zps5_ms_AMD}. 
Indeed, we have cross-checked that the four-loop result \eqref{eq:zps5_ms}, determined via direct calculations, can be fully reconstructed from \eqref{eq:zps5_ms_AMD} which only involves $\Zpsf$ up to three-loop order.
The explicit five-loop expression for $\Zpsbar$ reconstructed from \eqref{eq:zps5_ms_AMD} is provided in the supplementary material accompanying this article, along with all the aforementioned results for the reader's convenience.

\section{Conclusion}
\label{sec:conc}

In this work we have conducted a decomposition of the renormalization constant of the pseudoscalar operator defined with a non-anticommuting $\g5$ in dimensional regularization, in order to separate explicitly the component $\Zps5$ that arises solely from $\g5$-related symmetry-restoration, thereby facilitating the transformation to other (non-$\MSbar$) renormalization schemes. 
We have furthermore computed the complete four-loop result for the renormalization constant $\Zps5$ in QCD.
Additionally, by virtue of renormalization-group invariance, we have predicted the $\MSbar$ factor of the renormalization constant for this operator at the five-loop order in QCD.
We have also explained in general, how a result for $\Zpsbar$ at $N+1$-loop order can be efficiently derived only using an $N$-loop calculation for the pseudoscalar operator.

\section*{Acknowledgments}

The work of L.~C. was supported by the Natural Science Foundation of China under contract No.~12205171, No.~12235008, No.~12321005, and by the Taishan Scholar Foundation of Shandong province (tsqn202312052) and 2024HWYQ-005.
The work of M.~C.  was supported by the Deutsche Forschungsgemeinschaft (DFG) under grant 396021762 - TRR 257: Particle
Physics Phenomenology after the Higgs Discovery.

\bibliography{Zps5} 
\bibliographystyle{utphysM}
\end{document}